\documentclass{emulateapj}
\begin{document}
\slugcomment{Accepted for publication in the Astrophysical Journal Letters EVLA Special Edition}
 
\title{The Survey of HI in Extremely Low-mass Dwarfs (SHIELD)}

\author{John M. Cannon$^{1}$, 
Riccardo Giovanelli$^{2,3,4}$, 
Martha P. Haynes$^{2,3,4}$,
Steven Janowiecki$^{5,6}$,
Angela Parker$^{5,6}$,
John J. Salzer$^{5,6}$, 
Elizabeth A.K. Adams$^{2,3,4,6}$,
Eric Engstrom$^{1}$,  
Shan Huang$^{2,3,4}$,
Kristen B.W. McQuinn$^{7}$, 
J{\"u}rgen Ott$^{8,9}$, 
Am{\'e}lie Saintonge$^{10,11}$,
Evan D. Skillman$^{7}$,
John Allan$^{1}$, 
Grace Erny$^{1}$, 
Palmer Fliss$^{1}$, 
AnnaLeigh Smith$^{1}$}
\affil{\begin{scriptsize}
$^{1}$Department of Physics \& Astronomy, Macalester College, 1600 Grand Avenue, Saint Paul, MN 55105; jcannon@macalester.edu\\
$^{2}$Center for Radiophysics and Space Research, Space Sciences Bldg., Cornell University, Ithaca, NY 14853, USA\\
$^{3}$National Astronomy \& Ionosphere Center, Cornell University, Space Sciences Bldg., Ithaca, NY 14853, USA\\                                               
$^{4}$The National Astronomy \& Ionosphere Center is operated by Cornell University under a cooperative agreement with the National Science Foundation\\      
$^{5}$Department of Astronomy, Indiana University, 727 East Third Street, Bloomington, IN 47405, USA\\  
$^{6}$Visiting Astronomer, Kitt Peak National Observatory, National Optical Astronomy Observatories, which is operated by the Association of Universities for Research in Astronomy, Inc. (AURA) under cooperative agreement with the National Science Foundation. The WIYN Observatory is a joint facility of the University of Wisconsin-Madison, Indiana University, Yale University, and the National Optical Astronomy Observatories.\\
$^{7}$Astronomy Department, University of Minnesota, Minneapolis, MN 55455, USA\\   
$^{8}$National Radio Astronomy Observatory, P.O. Box O, Socorro, NM 87801, USA\\  
$^{9}$The National Radio Astronomy Observatory is operated by Associated Universities, Inc. under a cooperative agreement with the National Science Foundation\\  
$^{10}$Max Planck Institut f{\"u}r Astrophysik, Karl-Schwarzschildstrasse 1, D-85748 Garching, Germany\\  
$^{11}$Max Planck Institut f{\"u}r Extraterrestrische Physik, Giessenbachstrasse, D-85748 Garching, Germany
\end{scriptsize}}

\begin{abstract}

We present first results from the {\it Survey of HI in Extremely
  Low-mass Dwarfs} (SHIELD), a multi-configuration EVLA study of the
neutral gas contents and dynamics of galaxies with HI masses in the
10$^6$-10$^7$ M$_{\odot}$ range detected by the {\it Arecibo Legacy
  Fast ALFA} (ALFALFA) survey.  We describe the survey motivation and
concept demonstration using VLA imaging of 6 low-mass galaxies
detected in early ALFALFA data products.  We then describe the primary
scientific goals of SHIELD and present preliminary EVLA and WIYN~3.5m
imaging of the 12 SHIELD galaxies.  With only a few exceptions, the
neutral gas distributions of these extremely low-mass galaxies are
centrally concentrated.  In only 1 system have we detected HI column
densities higher than 10$^{21}$ cm$^{-2}$.  Despite this, the stellar
populations of all of these systems are dominated by blue stars.
Further, we find ongoing star formation as traced by H$\alpha$
emission in 10 of the 11 galaxies with H$\alpha$ imaging obtained to
date.  Taken together these results suggest that extremely low-mass
galaxies are forming stars in conditions different from those found in
more massive systems.  While detailed dynamical analysis requires the
completion of data acquisition, the most well-resolved system is
amenable to meaningful position-velocity analysis.  For AGC~749237, we
find well-ordered rotation of 30 km\,s$^{-1}$ at $\sim40\arcsec$
distance from the dynamical center. At the adopted distance of 3.2
Mpc, this implies the presence of a
{\raise0.3ex\hbox{$>$}\kern-0.75em{\lower0.65ex\hbox{$\sim$}}}1\,$\times$\,10$^8$
M$_{\odot}$ dark matter halo and a baryon fraction
{\raise0.3ex\hbox{$<$}\kern-0.75em{\lower0.65ex\hbox{$\sim$}}}0.1.

\end{abstract}						


\section{Populating the Cosmologically Important Faint End of the HI Mass Function:
ALFALFA}
\label{S1}

One of the major accomplishments of the {\it Arecibo Legacy Fast ALFA}
(ALFALFA) survey \citep{giovanelli05} is the detection of hundreds of
galaxies with HI masses $<$10$^8$ M$_{\odot}$.  As its design
intended, ALFALFA provides the first statistically robust sampling of
the faint end of the HI mass function \citep[M$_{\rm HI}$ $<$10$^8$
  M$_{\odot}$;][]{martin10}.  The sensitivity and resolution of
ALFALFA have now produced a collection of dozens of galaxies outside
the Local Group with HI masses below 10$^7$ M$_{\odot}$.  Each of
these objects has been cross-correlated with optical catalogs; {\it
  these extremely low-mass dwarfs are among the lowest-mass,
  gas-bearing systems that harbor detectable stellar populations in
  the local universe}.

Over the past few decades, the kinematics and stellar populations of
dwarf galaxies have been explored in detail, both in targeted
investigations and in dedicated surveys. These studies have focused
largely (though not exclusively) on systems that populate the HI mass
function above 10$^8$ M$_{\odot}$.  Three prominent surveys to this
end are the LITTLE THINGS \citep{hunter07}, FIGGS \citep{begum08}, and
VLA-ANGST \citep{ott10} programs.  While the selection and sample
criteria vary from one program to the next (including some overlap),
we note that the median HI masses are 8.5\,$\times$\,10$^7$
M$_{\odot}$, 2.7\,$\times$\,10$^7$ M$_{\odot}$, and
2.3\,$\times$\,10$^7$ M$_{\odot}$ for LITTLE THINGS, FIGGS, and
VLA-ANGST, respectively.  These surveys have revealed many new
insights into low-mass galaxies, including the characteristics of 21
systems (4, 8, and 9 in LITTLE THINGS, FIGGS, and VLA-ANGST,
respectively) with M$_{\rm HI} <$10$^7$ M$_{\odot}$.

\begin{deluxetable*}{cccccccccccc}
\tablecaption{VLA/EVLA Observations of ALFALFA-Selected Low-Mass Dwarf Galaxies}
\tablewidth{0pt}
\tablehead{
\colhead{AGC\tablenotemark{a}} &\colhead{$\alpha$}   &\colhead{$\delta$} &\colhead{Distance}   &\colhead{M$_{\rm r}$}       &\colhead{(u$-$r)}              &\colhead{M$_{\rm B}$} &\colhead{(B$-$V)}  &\colhead{\%Z$_{\odot}$\tablenotemark{b}} &\colhead{V$_{\rm 21}$}    &\colhead{W$_{\rm 21}$} &\colhead{log(M$_{\rm HI}$)}\\
\colhead{\#}  &\colhead{(J2000)}    &\colhead{(J2000)}  &\colhead{(Mpc)}      &\colhead{(mag)}            &{(mag)}                        &\colhead{(mag)}      &{(mag)}            &                         &\colhead{(km\,s$^{-1}$)} &\colhead{(km\,s$^{-1}$)} &\colhead{(M$_{\odot}$)}} 
\startdata
\multicolumn{12}{c}{Concept Demonstration Targets} \\ 
\cline{1-12}\\ 
100062                  &00:09:52.8 &15:43:58           &12.7\tablenotemark{c} &$-$15.02                  &1.02$\pm$0.02                  &N/A                     &N/A                   &N/A &869  &45 &7.82    \\ 
101772                  &00:11:08.2 &14:14:08           &11.7\tablenotemark{c} &$-$13.64                  &0.98$\pm$0.06                  &N/A                     &N/A                   &N/A &802  &37 &7.54    \\ 
111945                  &01:44:42.7 &27:17:18           &6.3\tablenotemark{c}  &$-$11.88\tablenotemark{d} &3.35$\pm$1.26\tablenotemark{d} &N/A                     &N/A                   &N/A &420  &38 &7.28        \\ 
321203                  &22:13:03.3 &28:04:28           &16.4\tablenotemark{c} &$-$14.20                  &0.98$\pm$0.04                  &N/A                     &N/A                   &14\% &983  &62 &7.83    \\  
321307                  &22:14:04.4 &25:41:08           &18.7\tablenotemark{c} &$-$13.88                  &1.24$\pm$0.07                  &N/A                     &N/A                   &N/A &1152 &60 &7.96     \\ 
332939                  &23:08:16.0 &31:53:57           &11.4\tablenotemark{c} &$-$13.67                  &0.69$\pm$0.05                  &N/A                     &N/A                   &N/A &692  &41 &7.74     \\ 
\cline{1-12}\\
\multicolumn{12}{c}{The SHIELD Sample} \\ 
\cline{1-12}\\
748778\tablenotemark{e}                  &00:06:34.3 &15:30:39           &5.4\tablenotemark{c}  &$-$10.52                  &0.81$\pm$0.19                  &$-$10.02             &0.25$\pm$0.03      &N/A   &258 &16 &6.51      \\
112521\tablenotemark{e}                  &01:41:07.6 &27:19:24           &7.2\tablenotemark{f}  &$-$11.52\tablenotemark{d} &2.19$\pm$0.42\tablenotemark{d} &$-$10.80             &0.45$\pm$0.03      &6\%   &274 &26 &6.92        \\
110482\tablenotemark{e}                  &01:42:17.4 &26:22:00           &7.2\tablenotemark{f}  &$-$13.63                  &1.25$\pm$0.04                  &$-$12.86             &0.49$\pm$0.02      &13\%  &357 &30 &7.21        \\
111946\tablenotemark{e}                  &01:46:42.2 &26:48:05           &7.2\tablenotemark{f}  &$-$11.49\tablenotemark{d} &1.46$\pm$0.25\tablenotemark{d} &$-$11.48             &0.39$\pm$0.03      &5\%   &367 &21 &6.97        \\
111977\tablenotemark{e}                  &01:55:20.2 &27:57:14           &5.5\tablenotemark{g}  &$-$12.55\tablenotemark{d} &2.26$\pm$0.16\tablenotemark{d} &$-$12.31             &0.47$\pm$0.02      &N/A   &207 &26 &6.78       \\
111164\tablenotemark{e}                  &02:00:10.1 &28:49:52           &4.9\tablenotemark{g}  &$-$11.50\tablenotemark{d} &0.80$\pm$0.09\tablenotemark{d} &$-$11.10             &0.41$\pm$0.02      &N/A   &163 &27 &6.57       \\
174585\tablenotemark{h,i}                &07:36:10.3 &09:59:11           &6.1\tablenotemark{c}  &N/A                       &N/A                            &N/A                  &N/A                &N/A   &356 &21 &6.68      \\
174605\tablenotemark{h}                  &07:50:21.7 &07:47:40           &6.0\tablenotemark{c}  &$-$10.46\tablenotemark{d} &1.85$\pm$0.21\tablenotemark{d} &$-$10.98             &0.47$\pm$0.05      &N/A   &351 &24 &6.75          \\
182595\tablenotemark{h}                  &08:51:12.1 &27:52:48           &6.1\tablenotemark{c}  &$-$12.45                  &1.27$\pm$0.05                  &$-$11.75             &0.52$\pm$0.05      &N/A   &398 &20 &6.66        \\
731457\tablenotemark{h}                  &10:31:55.8 &28:01:33           &5.4\tablenotemark{c}  &$-$12.55                  &1.23$\pm$0.03                  &$-$12.02             &0.39$\pm$0.05      &N/A   &454 &36 &6.63        \\
749237\tablenotemark{h}                  &12:26:23.4 &27:44:44           &3.2\tablenotemark{c}  &$-$11.58                  &1.28$\pm$0.03                  &$-$11.21             &0.44$\pm$0.05      &N/A   &372 &65 &6.64        \\
749241\tablenotemark{h}                  &12:40:01.7 &26:19:19           &4.3\tablenotemark{c}  &$-$9.27                   &0.83$\pm$0.15                  &$-$9.57              &0.22$\pm$0.05      &N/A   &451 &18 &6.52         
\enddata
\label{t1}
\tablenotetext{a}{Arecibo General Calalog}
\tablenotetext{b}{Assuming the Solar oxygen abundance From Asplund et\,al.\ (2009).}
\tablenotetext{c}{Derived using the parametric multiattractor flow model developed by \citet{masters05}; see further discussion in \citet{martin10}.}
\tablenotetext{d}{Magnitudes and colors uncertain due to {\it SDSS} shredding issues.}
\tablenotetext{e}{WIYN 3.5m imaging acquired in Fall 2010.}
\tablenotetext{f}{Probable member of NGC 672 group.}
\tablenotetext{g}{Tip of the red giant branch.}
\tablenotetext{h}{WIYN 3.5m imaging acquired in Spring 2011.}
\tablenotetext{i}{AGC 174585 is outside the SDSS footprint; an absolute calibration of the WIYN 3.5m observations awaits subsequent re-imaging.}
\end{deluxetable*}

The existence of such gas-rich galaxies with very shallow potential
wells poses interesting puzzles for the $\Lambda$CDM paradigm. Ram
pressure should easily strip their ISM via encounters with coronal gas
if they venture within the virial radius of nearby giant galaxies or
clusters \citep[e.g.,][]{lewis02,grebel03}. Star formation (SF)
activates feedback mechanisms that result in gas loss via superwinds;
simulations have predicted that galaxies with gas masses M$_{\rm HI}
<$10$^7$ M$_{\odot}$ are highly susceptible to mass loss via
starburst-driven superwinds \citep[e.g.,][]{maclow99,ferrara00}, and
observational results support this scenario
\citep[e.g.,][]{martin02,ott05}.  The metagalactic UV radiation field
inhibits gas accretion and cooling most severely in low-mass halos
\citep{rees86,babul92,benson02,hoeft06}.  A hot IGM should vaporize a small,
unshielded cold gas mass within less than a Hubble time
\citep{benson02}.  And yet, despite these myriad destruction
mechanisms, objects such as Leo T (M$_{\rm HI}$ $\simeq$10$^5$
M$_{\odot}$, SF within the last few hundred Myr; Irwin et\,al.\ 2007)
exist. The survival of such systems seems to depend on both the
protection provided by a shielding envelope of warm, ionized gas
\citep{sternberg02}, and on environment i.e., they are found in
locations removed from the immediate vicinity of dense coronal
gas. For halo masses below $\sim10^9$ M$_{\odot}$, the baryon fraction
($f_b$) of dwarf systems is predicted to fall below the cosmic value
of $f_b$$\sim$0.16 \citep{hoeft06,mcgaugh10}.  Most of the remaining
baryons within a given system are in the form of a warm, ionized phase
that is difficult to detect.  In the central regions of these halos, a
small fraction of the gas is predicted to be neutral and thus capable
of condensing into stars.

The extremely low-mass galaxies observed in the local universe by
ALFALFA have not been detected in large enough numbers to solve the
``missing satellite problem'' \citep{klypin99}.  However, these
systems are nonetheless critical for models of cosmological structure
formation for two primary reasons.  First, an understanding of their
characteristics as the survivors of the structure formation process
provides constraints on the aforementioned destruction mechanisms
acting on low-mass halos.  Second, these systems bridge the
observational gap between the well-studied and comparatively high-mass
(10$^7$ M$_{\odot}$ $<$ M$_{\rm HI}$ $<$ 10$^9$ M$_{\odot}$) dwarfs of
the local universe and the extremely low-mass ultra-faint dwarf
companions of the Milky Way (e.g., Willman~1, with M$_{\rm
  HI}\simeq$5\,$\times$\,10$^5$ M$_{\odot}$; {Willman
  et\,al. 2005}\nocite{willman05}; {Martin
  et\,al. 2007}\nocite{martin07}).  To probe this transition, we have
thus undertaken a multiwavelength study of very low-mass (M$_{\rm HI}$
{\raise0.3ex\hbox{$<$}\kern-0.75em{\lower0.65ex\hbox{$\sim$}}}10$^7$
M$_{\odot}$) dwarfs across the range of halo masses over which the
transition from $f_b\simeq 0.16$ to $<0.01$ takes place
\citep[see][]{hoeft06,mcgaugh10}.  This {\it Letter} presents first
results from optical and radio imaging of extremely low-mass galaxies
that form part of this observing campaign.

\begin{figure*}
\plotone{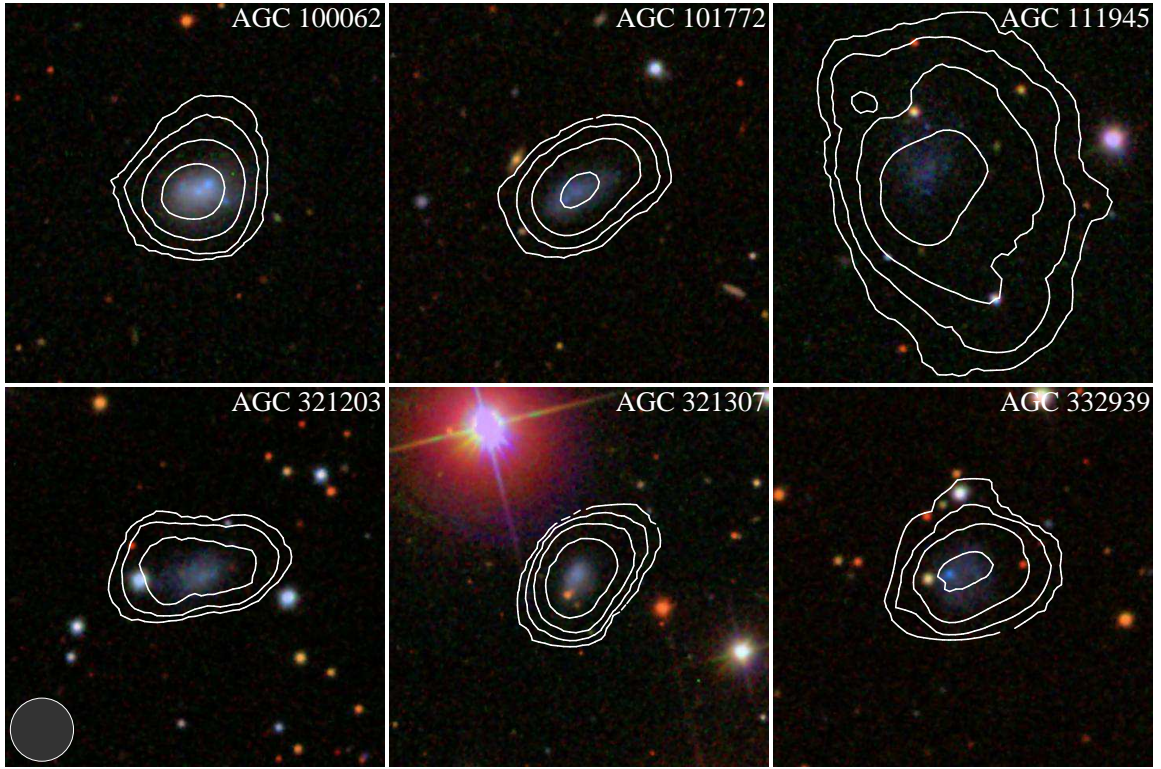}
\caption{{\it SDSS} 3-color images of the precursor galaxies, with VLA
  HI column density contours overlaid at the (0.5, 1, 2,
  4)\,$\times$\,10$^{20}$ cm$^{-2}$ levels. Each panel is
  2\arcmin\,$\times$\,2\arcmin\ square, with north up and east to the
  left; the HI images all have a 20\arcsec\ beam size (shown in the
  bottom left panel). Note the visually blue stellar populations and
  low surface density neutral gas in these systems.}
\label{figcap1}
\end{figure*}

\section{Exploring the Faint End of the HI Mass Function: SHIELD}
\label{S2}

The measurement of the total halo masses of the extremely low-mass
galaxies detected by ALFALFA requires deep, high spatial and spectral
resolution HI observations in order to characterize the systems'
rotation curves.  This measurement is the primary motivation for {\it
  The Survey of HI in Extremely Low-mass Dwarfs} (SHIELD).  We now
describe the strategy, goals, and present status of SHIELD.

Our investigation began with a series of concept demonstration
observations of 6 low-mass systems detected in early ALFALFA
observations (see ``Concept Demonstration Targets'' in
Table~\ref{t1}).  These systems were observed with the {\it Very Large
  Array\footnote{The National Radio Astronomy Observatory (NRAO) is a
    facility of the National Science Foundation operated under
    cooperative agreement by Associated Universities, Inc.}}  (VLA)
for programs AC963 (PI Cannon) and AS883 (PI Saintonge) during the
VLA\,$\rightarrow$\,EVLA transition.  The data reduction and imaging
were performed wtih the Astronomical Image Processing System
(AIPS\footnote{The Astronomical Image Processing System (AIPS) has
  been developed by the NRAO.}); reductions proceeded normally, with
modifications as necessary for aliased baselines.  The calibrated
datasets were imaged to the rms level using both natural (ROBUST=5)
and robust (ROBUST=0.5) weighting, and then smoothed to a common
circular beam size of 20\arcsec.  The two approaches resulted in flux
integrals that differed by
{\raise0.3ex\hbox{$<$}\kern-0.75em{\lower0.65ex\hbox{$\sim$}}}10\% for
most systems, and we thus show only the robust weighted images in this
work.

Figure~\ref{figcap1} shows {\it SDSS} 3-color images of the six
precursor galaxies.  It is immediately obvious that the high surface
brightness stellar populations are dominated by blue stars.  This is
confirmed by the {\it SDSS} photometry presented in Table~\ref{t1}.
The weighted mean (u$-$r) color for the six precursor systems is
(u$-$r) $=$ 0.99\,$\pm$\,0.02.  This can be compared with
(u$-$r)$\simeq$1.3, the mean SDSS color for all ALFALFA galaxies with
log(M$_{\rm HI}$)$<$7.7 (Huang et\,al. 2011, in preparation).

Figure~\ref{figcap1} compares the stellar and neutral gas
distributions; overlaid on each panel are HI column density contours
at the (0.5, 1, 2, 4)\,$\times$\,10$^{20}$ cm$^{-2}$ levels.  The
neutral gas distributions are centrally concentrated and spatially
coincident with the blue stellar populations.  All systems show peak
column densities below the canonical SF surface density threshold of
n$_{\rm HI} = $ 10$^{21}$ cm$^{-2}$ \citep{skillman87,kennicutt98}.
While the spectral resolution of these images are sufficient to
resolve the HI profiles derived from ALFALFA, the HI distributions are
only slightly larger than the synthesized beam size (20\arcsec).
Although these six systems have HI masses above 10$^7$ M$_{\odot}$
(see Table~\ref{t1}), these precursor observations nonetheless
demonstrate the scientific potential of a coordinated observing
campaign using the EVLA (see {Perley et\,al. 2011}\nocite{perley11}
for a description of the EVLA project) to explore the low-mass
galaxies detected by ALFALFA.

\begin{figure*}
\plotone{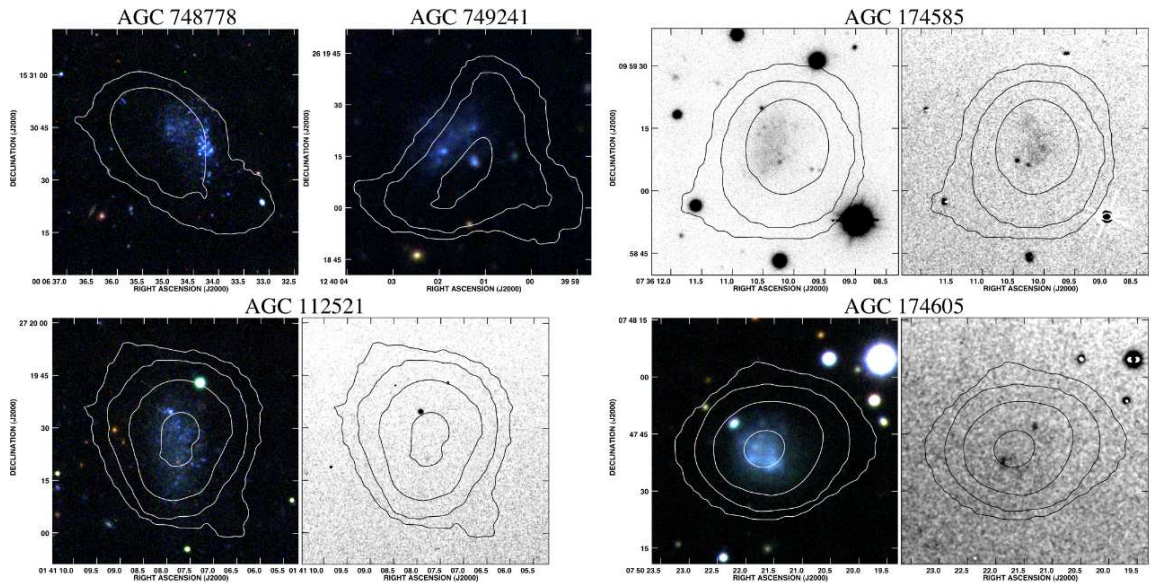}
\caption{WIYN 3.5m 3-color (BVR) and continuum-subtracted H$\alpha$
  images of 5 SHIELD galaxies; AGC 748778 is an H$\alpha$
  non-detection, while cloud cover compromised the H$\alpha$ image of
  AGC 749241 and the B and V images of AGC 174585 (hence only the
  R-band image is shown).  EVLA HI column density contours are
  overlaid at the (0.5, 1, 2, 4, 8)\,$\times$\,10$^{20}$ cm$^{-2}$
  levels.  The beam size is 20\arcsec\ in each panel; note that the
  fields of view are tailored to each galaxy to display the most
  detail in the stellar populations, and thus the beam sizes are best
  assessed using the coordinate axes.  Note the remarkably blue
  stellar populations, low surface density neutral gas, and ongoing SF
  in each of these systems.}
\label{figcap2}
\end{figure*}

We thus initiated SHIELD (OSRO Program 10B-187, PI Cannon; time
allocation of 180 hours), a multi-configuration study of the neutral
ISM of 12 extremely low-mass systems.  The sample members were
selected from the $>$11,000 ALFALFA-detected galaxies to date on the
basis of HI mass (M$_{\rm HI}$ $\le$1.6\,$\times$\,10$^7$ M$_{\odot}$)
and line width (full width at 50\% of peak $<$ 65 km\,s$^{-1}$ from
ALFALFA); the latter discriminates against massive but HI-poor
galaxies and identifies the truly low mass galaxies.  The median
distance, HI mass, and HI line width are 5.7 Mpc,
4.7\,$\times$\,10$^6$ M$_{\odot}$, and 25 km\,s$^{-1}$, respectively.
Our observational strategy (9, 4, 2 hours per source in the B, C, D
arrays, respectively, with typical calibration overheads of 25\%)
achieves high spatial ($\sim$6\arcsec\ synthesized beam at full
resolution) and spectral resolution ($\sim$0.82
km\,s$^{-1}$\,ch$^{-1}$), while retaining sensitivity to extended
structure.  When data acquisition is complete, the 5\,$\sigma$ (per
channel) column density sensitivity will be n$_{\rm HI}$ $>$10$^{19}$
cm$^{-2}$ and $>$2.3\,$\times$\,10$^{20}$ cm$^{-2}$ at low and high
spatial resolution, respectively; the realized column density
sensitivities will be higher because the linewidths are larger than
the channel spacing (see Table~\ref{t1}).  The WIDAR correlator is
used to provide a single 1\,MHz sub-band with 2 polarization products
and 256 channels each, covering 211 km\,s$^{-1}$ of frequency space at
3.906 kHz\,ch$^{-1}$.  Data acquisition for SHIELD began in October
2010.  The images presented here use only the data acquired in the C
configuration; the reduction techniques are standard, and the same
imaging techniques are used for the SHIELD galaxies as described above
for the precursor galaxies.  The rms noise values in the robust
weighted cubes range from 0.8-1.0 mJy\,Bm$^{-1}$.

\begin{figure*}
\plotone{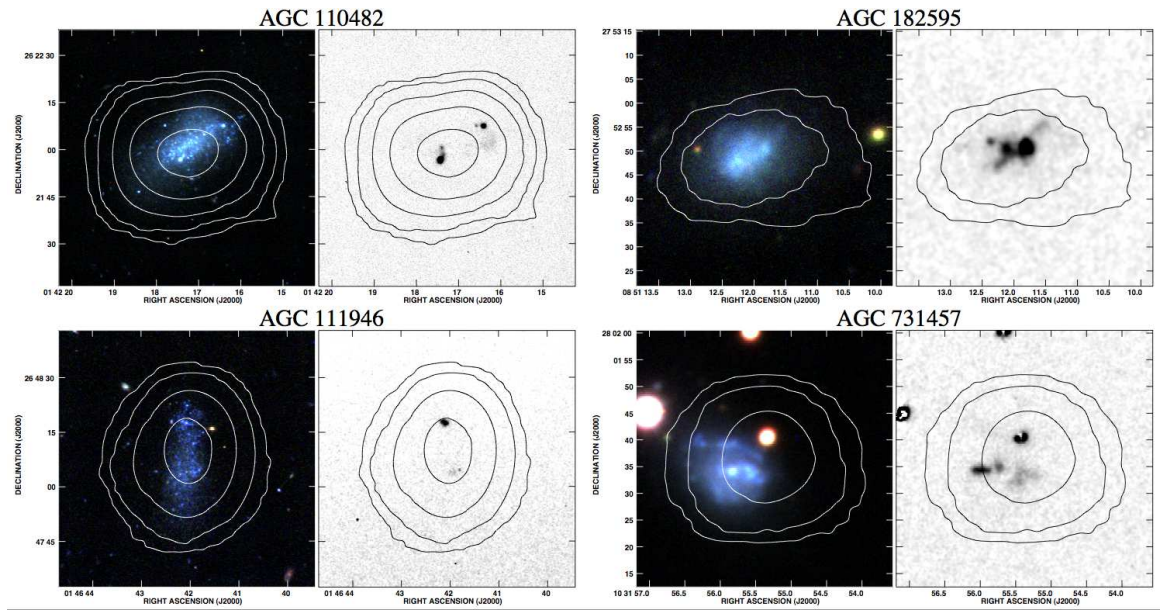}
\caption{Same as Figure~\ref{figcap2}, for 4 more SHIELD galaxies.}
\label{figcap3}
\end{figure*}

Figures~\ref{figcap2}, \ref{figcap3} and \ref{figcap4} present
comparisons of the optical and HI properties of all 12 galaxies using
the C-configuration datasets only.  The optical images were acquired
with the WIYN\footnote{The WIYN Observatory is a joint facility of the
  University of Wisconsin-Madison, Indiana University, Yale
  University, and the National Optical Astronomy Observatories.}  3.5m
telescope and the Mini-Mosaic camera during two observing runs
(October 2010 and March 2011).  The Fall images (see Table~\ref{t1})
were obtained during superior observing conditions
($\sim$0.4\arcsec\ FWHM seeing); the Spring images (see
Table~\ref{t1}) were obtained during average observing conditions
({\raise0.3ex\hbox{$>$}\kern-0.75em{\lower0.65ex\hbox{$\sim$}}}1\arcsec\ FWHM
seeing). Broadband Johnson-Cousins (B, V, R) filters, and a narrowband
H$\alpha$ filter, were used.  Standard reduction strategies were
applied.  The images are first cosmic ray rejected, aligned and
combined.  The broad and narrowband images are then smoothed to a
common point spread function size and flux scaled to remove the
continuum from the narrowband image.  Finally, standard photometry
routines are applied, using observations of photometric standard stars
acquired during the same observing session.  We estimate the
photometric accuracies of the Fall and Spring images to be better than
2\% and 4\%, respectively.

Table~\ref{t1} compiles the absolute B magnitudes and (B$-$V) colors
of 11 of the 12 SHIELD galaxies; an absolute calibration of the images
of AGC\,174585 (which is outside the SDSS footprint) was not attained.
Similar to the precursor systems shown in Figure~\ref{figcap1}, the
optical appearances of the 12 galaxies shown in Figures~\ref{figcap2},
\ref{figcap3} and \ref{figcap4} are each dominated by blue stars (note
that only the R-band image of AGC 174585 is shown, due to two of the
images being affected by cloud cover).  The weighted mean (B$-$V)
color is 0.43\,$\pm$\,0.01; this compares well with the typical
(B$-$V) $=$0.42\,$\pm$\,0.05 found for local star-forming dIrr
galaxies \citep{vanzee97,vanzee00}.  Similarly, the weighted mean
(u$-$r) color for the SHIELD galaxies is 1.25\,$\pm$\,0.02; these
systems are on average slightly redder than the precursor objects,
though still consistent with the results of Huang et\,al. (in
preparation).

SHIELD is a systematic investigation of a sample of galaxies with HI
masses below 10$^7$ M$_{\odot}$ outside the Local Group.  Using our
EVLA datasets, we are focusing on three primary goals.  First, what
properties change between mini-halos (HI clouds without optical
counterparts; see {Giovanelli et\,al. 2010}\nocite{giovanelli10}),
very low-mass dwarfs, and more massive systems?  Are the
cosmologically important galaxies with HI masses of 10$^6$--10$^7$
M$_{\odot}$ systematically different than more or less massive
objects?  We plan to identify correlations between fundamental galaxy
parameters (e.g., HI line width, stellar mass, SF rate, etc.) by
undertaking a comparative study of these properties in objects that
span some three orders of magnitude in dynamical mass, from
$\sim$10$^6$--10$^9$ M$_{\odot}$ (i.e., covering the mini-halo to
dwarf galaxy transition region).  The synthesis observations provided
by the EVLA are crucial components of this program, as they allow us
to examine both the global and the local processes that influence the
evolution of these systems.

Second, what fraction of the mass in these low-mass dwarfs is
baryonic? When data acquisition is complete, our EVLA observing
campaign will allow us to study the ISM kinematics on spatial scales
of order 200 pc in the galaxies in our sample (assuming a
6\arcsec\ beam and a maximum distance of $\sim$8 Mpc).  These data
will allow us to extract detailed rotation curves in the inner regions
of these galaxies and to infer the dark-to-baryonic fraction
throughout the disks.  Equally important is the sensitivity of the
combined EVLA\,$+$\,Arecibo datasets to extended structure, which we
will exploit to study gas in the outermost regions of the disks.

Third, is the character of the SF process different in very low-mass
galaxies?  These systems have retained HI mass reservoirs of
10$^6$-10$^7$ M$_{\odot}$ over a Hubble time; they are apparently
evolving in relative quiescence and are exceedingly inefficient at
converting their interstellar gas into stars.  These clues suggest
that the SF law may in fact deviate significantly from the canonical
Schmidt-Kennicutt prescription derived for more massive systems
\citep{skillman87,kennicutt98}.  

\section{Preliminary Results}
\label{S3}

While data acquisition for SHIELD is ongoing, we are able to draw
conclusions about various intriguing properties of these extremely
low-mass galaxies using the C configuration images alone.  We discuss
three results in turn below.  First, with only one exception
(AGC\,111977), the HI distributions of the SHIELD galaxies are
centrally concentrated at 20\arcsec\ resolution.  In only one case
(AGC\,110482) does the observed HI column density reach the 10$^{21}$
cm$^{-2}$ level; however, we expect that higher-resolution imaging
will localize regions of larger column densities (see, e.g., the
discussion of beam smearing effects in {Begum
  et\,al. 2008}\nocite{begum08}).  The SHIELD galaxies appear to have
gas surface densities comparable to those in other nearby galaxies
with similar HI masses (e.g., all of the galaxies with M$_{\rm HI} <$
10$^7$ M$_{\odot}$ in {Begum et\,al. 2008}\nocite{begum08} have peak
n$_{\rm HI} < $ 10$^{21}$ cm$^{-2}$).

\begin{figure*}
\epsscale{0.9}
\plotone{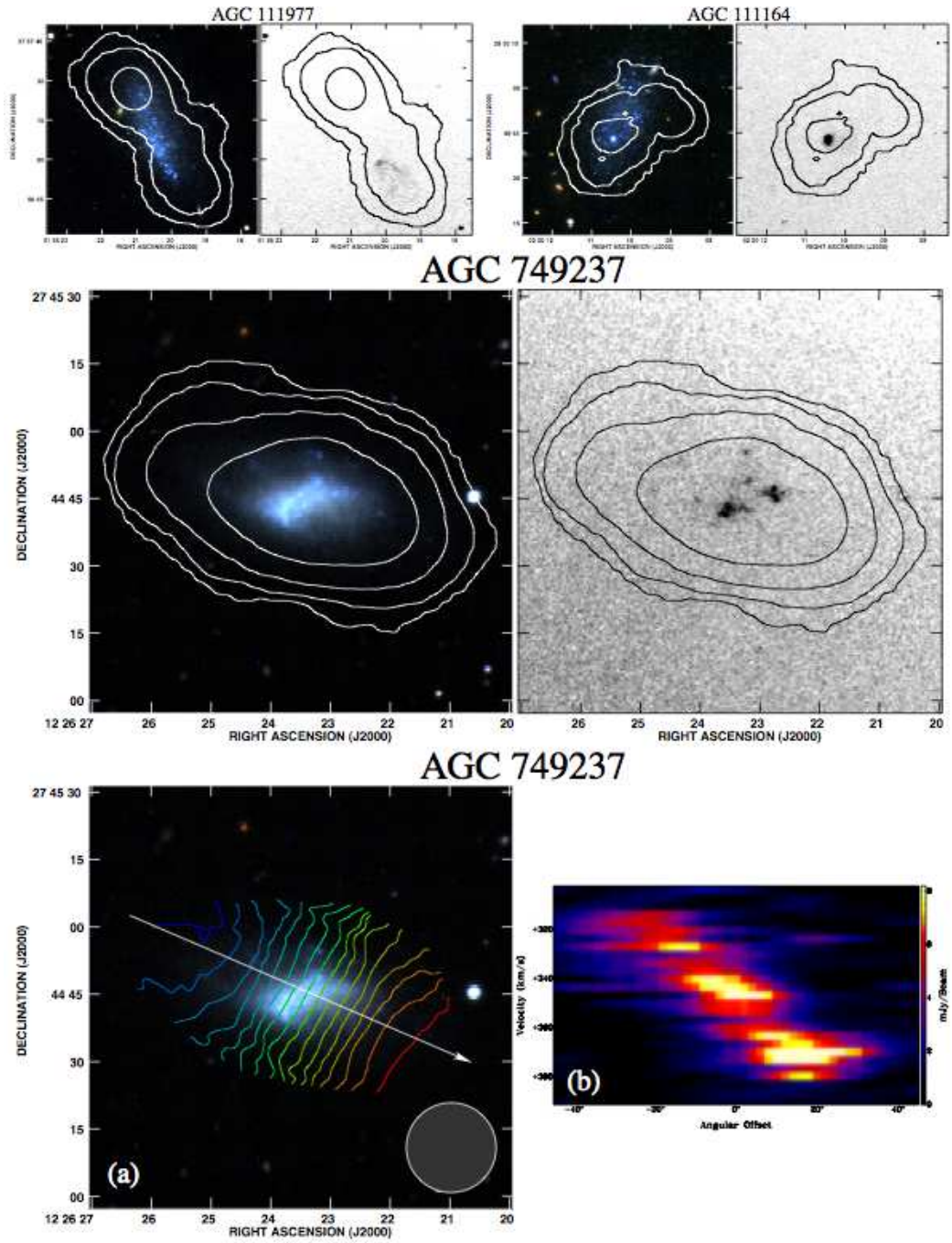}
\epsscale{1.0}
\caption{Same as Figure~\ref{figcap2}, for 3 more SHIELD galaxies.
  The bottom panels show the HI distribution and kinematics of the
  most highly-resolved system, AGC 749237. (a) shows color-coded
  isovelocity contours between 325 and 370 km\,s$^{-1}$ in intervals
  of 3 km\,s$^{-1}$ per contour. The position of the HI major-axis
  position-velocity cut (position angle 246.7{\hbox{$^\circ$}},
  measured east of north) is indicated by the white arrow.  The
  20\arcsec\ beam size is shown at the lower right. (b) shows the
  resulting position-velocity diagram; the rotation is almost
  perfectly solid-body over the inner 40\arcsec\ (620 pc at the
  adopted distance of 3.2 Mpc).}
\label{figcap4}
\end{figure*}

Second, as discussed in detail by \citet{begum06} and Roychowdhury
et\,al. ({2009}\nocite{roychowdhury09},
{2011}\nocite{roychowdhury11}), the relationship between HI column
density and ongoing SF in low-mass galaxies is complex.  Stochastic
effects play an important role, and the overall process appears to be
less efficient than in more massive galaxies.  When taken at face
value, the low integrated column densities of the SHIELD galaxies are
not conducive to ongoing SF.  And yet, the stellar populations of
these systems are dominated by blue stars (as evidenced by the blue
integrated colors in Table~\ref{t1}), and the majority of the galaxies
show active SF as traced by H$\alpha$ emission (see
Figures~\ref{figcap2}, \ref{figcap3} and \ref{figcap4}).  Our
WIYN~3.5m imaging produced usable H$\alpha$ images of 11 systems
(images of AGC 749241 were acquired but were compromised by cloud
cover).  10 of these systems show high surface brightness H$\alpha$
emission; the only conclusive H$\alpha$ non-detection is AGC
748778. In most cases, H$\alpha$ emission is located in close
proximity to the highest HI column densities (but see the diffuse
H$\alpha$ emission in the southern region of AGC 111977); however, as
noted above, none of these columns exceeds the canonical SF surface
density threshold of 10$^{21}$ cm$^{-2}$.  We note with interest that
two of the six precursor systems (AGC 100062 and AGC 101772) have
strong H$\alpha$ emission lines in their {\it SDSS} spectra.

Finally, the full SHIELD datasets will allow us to test our
understanding of the dark matter contents of these low-mass galaxies
and of $f_b$ within them. While detailed rotation curve analysis
requires higher resolution B-configuration EVLA images, a few systems
are already amenable to meaningful position-velocity analysis using
the 20$\arcsec$ datacubes alone. Using AGC\,749237 as an example (see
the velocity field and position-velocity slice in
Figure~\ref{figcap4}), we find solid-body rotation of 30 km\,s$^{-1}$
at a distance of $\sim$40$\arcsec$ from the dynamical center. We apply
no inclination correction to this velocity; we note, however, that the
optical inclination {\it i} $\simeq$56{\hbox{$^\circ$}}, and thus the
true rotational velocity may be higher by $\sim$20\% or more.  At the
adopted (but uncertain) distance of 3.2 Mpc, this implies a dynamical
mass M$_{\rm dyn}$
{\raise0.3ex\hbox{$>$}\kern-0.75em{\lower0.65ex\hbox{$\sim$}}}
$\sim$1\,$\times$\,10$^8$ M$_{\odot}$.  Correcting for helium and
molecular gas (35\% of M$_{\rm HI}$) and assuming equal masses of
stars and gas, this implies $f_b$
{\raise0.3ex\hbox{$<$}\kern-0.75em{\lower0.65ex\hbox{$\sim$}}}0.1. As
expected, this low-mass galaxy follows the trends of $f_b$ at low
dynamical masses in the models of \citet{hoeft06} and
\citet{mcgaugh10}.

\section{Conclusion}
\label{S4}

We have introduced SHIELD, a systematic investigation of a sample of
galaxies with HI masses below 10$^7$ M$_{\odot}$ outside the Local
Group.  A primary goal of SHIELD and of ALFALFA is to characterize
changes in fundamental galaxy properties as functions of total halo
mass.  The EVLA imaging described in this {\it Letter} is the
centerpiece of a multiwavelength observing campaign designed to place
these low-mass systems in an evolutionary context.  When completed,
the SHIELD data suite will offer a unique opportunity to study the
fundamental properties of galaxies in a newly-opened region of
parameter space.

\acknowledgements
 
The authors would like to acknowledge the work of the entire ALFALFA
collaboration team in observing, flagging, and extracting the catalog
of galaxies used to identify the SHIELD sample. The ALFALFA team at
Cornell is supported by NSF grant AST-0607007 to R.G. and M.P.H. and
by a grant to M.P.H. from the Brinson Foundation. E.A.K.A. is
supported by a NSF predoctoral fellowship.  J.M.C. is grateful to the
NRAO for supporting a very productive and enjoyable sabbatical leave,
during which this manuscript was written.  J.M.C. thanks Eric Greisen,
Emmanuel Momjian, and Gustaaf Van Moorsel for helpful discussions, and
Macalester College for research and teaching support.

Funding for the SDSS and SDSS-II has been provided by the Alfred
P. Sloan Foundation, the Participating Institutions, the National
Science Foundation, the U.S. Department of Energy, the National
Aeronautics and Space Administration, the Japanese Monbukagakusho, the
Max Planck Society, and the Higher Education Funding Council for
England. The SDSS Web Site is http://www.sdss.org/.

The SDSS is managed by the Astrophysical Research Consortium for the
Participating Institutions. The Participating Institutions are the
American Museum of Natural History, Astrophysical Institute Potsdam,
University of Basel, University of Cambridge, Case Western Reserve
University, University of Chicago, Drexel University, Fermilab, the
Institute for Advanced Study, the Japan Participation Group, Johns
Hopkins University, the Joint Institute for Nuclear Astrophysics, the
Kavli Institute for Particle Astrophysics and Cosmology, the Korean
Scientist Group, the Chinese Academy of Sciences (LAMOST), Los Alamos
National Laboratory, the Max-Planck-Institute for Astronomy (MPIA),
the Max-Planck-Institute for Astrophysics (MPA), New Mexico State
University, Ohio State University, University of Pittsburgh,
University of Portsmouth, Princeton University, the United States
Naval Observatory, and the University of Washington.

\bibliographystyle{apj}                                                 

\begin{thebibliography}{}

\bibitem[Babul \& Rees(1992)]{babul92} Babul, A., \& Rees,
  M.~J.\ 1992, \mnras, 255, 346

\bibitem[Begum et\,al.(2006)]{begum06} Begum, A., Chengalur, J.~N.,
  Karachentsev, I.~D., Kaisin, S.~S., \& Sharina, M.~E.\ 2006, \mnras,
  365, 1220

\bibitem[Begum et\,al.(2008)]{begum08} Begum, A., Chengalur, J.~N.,
  Karachentsev, I.~D., Sharina, M.~E., \& Kaisin, S.~S.\ 2008, \mnras,
  386, 1667

\bibitem[Benson et\,al.(2002)]{benson02} Benson, A.~J., Lacey, C.~G.,
  Baugh, C.~M., Cole, S., \& Frenk, C.~S.\ 2002, \mnras, 333, 156

\bibitem[Ferrara \& Tolstoy(2000)]{ferrara00} Ferrara, A., \& Tolstoy,
  E.\ 2000, \mnras, 313, 291

\bibitem[Giovanelli et\,al.(2005)]{giovanelli05} Giovanelli, R., et\,al.\ 2005, \aj, 130, 2598

\bibitem[Giovanelli et\,al.(2010)]{giovanelli10} Giovanelli, R., 
Haynes, M.~P., Kent, B.~R., \& Adams, E.~A.~K.\ 2010, \apjl, 708, L22 

\bibitem[Grebel et\,al.(2003)]{grebel03} Grebel, E.~K., Gallagher,
  J.~S., III, \& Harbeck, D.\ 2003, \aj, 125, 1926

\bibitem[Hoeft et\,al.(2006)]{hoeft06} Hoeft, M., Yepes, G., 
Gottl{\"o}ber, S., \& Springel, V.\ 2006, \mnras, 371, 401

\bibitem[Hunter et\,al.(2007)]{hunter07} Hunter, D.~A., Brinks, 
E., Elmegreen, B., Rupen, M., Simpson, C., Walter, F., Westpfahl, D., 
\& Young, L.\ 2007, Bulletin of the American Astronomical Society, 38, 895

\bibitem[Kennicutt(1998)]{kennicutt98} Kennicutt, R.~C., Jr.\ 1998,
  \apj, 498, 541

\bibitem[Klypin et\,al.(1999)]{klypin99} Klypin, A., Kravtsov, 
A.~V., Valenzuela, O., \& Prada, F.\ 1999, \apj, 522, 82 

\bibitem[Lewis et\,al.(2002)]{lewis02} Lewis, I., et\,al.\ 2002,
  \mnras, 334, 673

\bibitem[Mac Low \& Ferrara(1999)]{maclow99} Mac Low, M.-M., \&
  Ferrara, A.\ 1999, \apj, 513, 142

\bibitem[Martin et\,al.(2002)]{martin02} Martin, C.~L., Kobulnicky,
  H.~A., \& Heckman, T.~M.\ 2002, \apj, 574, 663

\bibitem[Martin et al.(2007)]{martin07} Martin, N.~F., Ibata, R.~A.,
  Chapman, S.~C., Irwin, M., \& Lewis, G.~F.\ 2007, \mnras, 380, 281

\bibitem[Martin et\,al.(2010)]{martin10} Martin, A.M., Papastergis,
  E., Giovanelli, R., Haynes, M.P., Springob, C.M.  \& Stierwalt,
  S. 2010, \apj, 723, 1359

\bibitem[Masters(2005)]{masters05} Masters, K.~L.\ 2005, 
Ph.D.~Thesis, Cornell University

\bibitem[McGaugh et\,al.(2010)]{mcgaugh10} McGaugh, S.~S., Schombert,
  J.~M., de Blok, W.~J.~G., \& Zagursky, M.~J.\ 2010, \apjl, 708, L14

\bibitem[Ott et\,al.(2005)]{ott05} Ott, J., Walter, F., \& Brinks,
  E.\ 2005, \mnras, 358, 1453

\bibitem[Ott et\,al.(2010)]{ott10} Ott, J., Warren, S., Stilp, A.,
  Skillman, E., Dalcanton, J., Walter, F., \& de Blok, E.\ 2010,
  Bulletin of the American Astronomical Society, 42, \#202.03

\bibitem[Perley et\,al.(2011)]{perley11} Perley, R.~A., Chandler, C.~J.,
  Butler, B.~J., \& Wrobel, J.~M.\ 2011, \apjl, in press

\bibitem[Rees(1986)]{rees86} Rees, M.~J.\ 1986, \mnras, 218, 25P

\bibitem[Roychowdhury et\,al.(2009)]{roychowdhury09} Roychowdhury, S., 
Chengalur, J.~N., Begum, A., 
\& Karachentsev, I.~D.\ 2009, \mnras, 397, 1435

\bibitem[Roychowdhury et\,al.(2011)]{roychowdhury11} Roychowdhury, S.,
  Chengalur, J.~N., Kaisin, S.~S., Begum, A., \& Karachentsev,
  I.~D.\ 2011, \mnras, L256

\bibitem[Skillman(1987)]{skillman87} Skillman, E.~D.\ 1987, NASA
  Conference Publication, 2466, 263

\bibitem[Sternberg et\,al.(2002)]{sternberg02} Sternberg, A., McKee,
  C.~F., \& Wolfire, M.~G.\ 2002, \apjs, 143, 419

\bibitem[van Zee et\,al.(1997)]{vanzee97} van Zee, L., Haynes, M.~P.,
  \& Salzer, J.~J.\ 1997, \aj, 114, 2479

\bibitem[van Zee(2000)]{vanzee00} van Zee, L.\ 2000, \aj, 119, 2757

\bibitem[Willman et\,al.(2005)]{willman05} Willman, B., et al.\ 
2005, \apjl, 626, L85 


\end{thebibliography}

\end{document}